\documentclass[
    ,final            % use final for the camera ready runs
  ]
  {aipproc}
\usepackage{amsmath,amssymb}
\layoutstyle{8x11double}

\begin{document}

\title{Band Engineering in Cooper-Pair Box: Dispersive Measurements of Charge and Phase}



\author{Mika Sillanp\"a\"a\footnote{present address: National Institute of Standards and Technology, 325 Broadway, Boulder, CO 80305, USA}, Leif Roschier, Teijo Lehtinen and Pertti Hakonen}
{address={Low Temperature Laboratory, Helsinki University of
Technology, FIN-02015 HUT, Finland}}
\begin{abstract}
Low-frequency susceptibility of the split Cooper-pair box (SCPB)
is investigated for use in sensitive measurements of external
phase or charge. Depending on the coupling scheme, the box appears
as either inductive or capacitive reactance which depends on
external phase and charge. While coupling to the source-drain
phase, we review how the SCPB looks like a tunable inductance,
which property we used to build a novel radio-frequency
electrometer. In the dual mode of operation, that is, while
observed at the gate input, the SCPB looks like a capacitance. We
concentrate on discussing the latter scheme, and we show how to do
studies of fast phase fluctuations at a sensitivity of 1
mrad/$\sqrt{\mathrm{Hz}}$ by measuring the input capacitance of
the box.
\end{abstract}

\maketitle

\section{Introduction}

Josephson junctions (JJ) store energy according to $E = -E_J \cos
(\varphi)$, where $\varphi$ is the phase difference across the
junction, and the Josephson energy $E_J$ is related to the
junction critical current $I_C$ through $I_C = 2e E_J /\hbar$.
Since JJ's also typically exhibit negligible dissipation, they can
be used as reactive circuit components. By combining the Josephson
equations $I = I_C \sin (\varphi)$ and $\dot{\varphi} = 2e V(t) /
\hbar$, where $V(t)$ is the voltage across the junction, we find
that a single JJ behaves as a nonlinear inductance,
\begin{equation}\label{eq:LJNonLin}
    L_J (\varphi) = \frac{\hbar}{2 e I_C \cos (\varphi)} = \frac{L_{J0}}{\cos
    (\varphi)},
\end{equation}
where we defined the linear-regime Josephson inductance $L_{J0} =
\hbar /(2 e I_C)$.

\begin{figure}[h]
  \includegraphics[width=7.5cm]{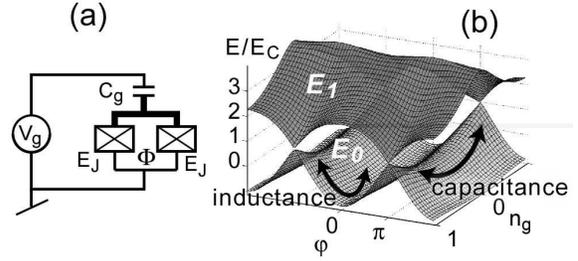}\label{fig:EnergySurface}
\caption{(a) Schematics of the SCPB. The mesoscopic island (thick
line) has a total capacitance $C_{\Sigma}$ and charging energy
$E_C = e^2/(2 C_{\Sigma})$; (b) two lowest energy bands $E_k$ ($k
= 0, 1$) of the SCPB , for $E_J/E_C = 1.7$ (without the parabolic
background $- (n_g e)^2/(2 C_g)$, see Eq.~(\ref{eq:bands})).
Inductive and capacitive susceptibilities are illustrated by the
arrows parallel to $\varphi$ and $n_g$, respectively.}
\end{figure}

Quantum effects in mesoscopic JJ's \cite{widom84,Likharev85} may
modify Eq.~(\ref{eq:LJNonLin}) in an important manner. In
particular, the Josephson reactance may become capacitive
\cite{AverinBruder03,Leif05}. In this brief communication, we
investigate the Josephson reactance in the split Cooper-pair box
(SCPB) geometry, with emphasis on detector applications. We first
review the inductive susceptibility, and then concentrate on
discussing the capacitive susceptibility in the spirit of a novel
phase detector. The discussion relies heavily on the energy bands
\cite{lukens} $E_k$ of the SCPB, two lowest of them given in the
limit $E_J/E_C \ll 1$ as
\begin{equation}\label{eq:bands}
\begin{split}
E_{0,1} & =  E_C (n_g^2 - 2 n_g + 2) \mp \\
& \sqrt{\left(E_J \cos(\varphi/2) \right)^2 + \left(2 E_C (1-n_g)
\right)^2} - C_g V_g^2/2
\end{split}
\end{equation}
as a function of the classical fields $\varphi = 2\pi \Phi /
\Phi_0$ and $n_g = C_g V_g / e$ (see
Fig.~\ref{fig:EnergySurface}).

\section{Quantum inductance}

\begin{figure}[h]
  \includegraphics[width=7cm]{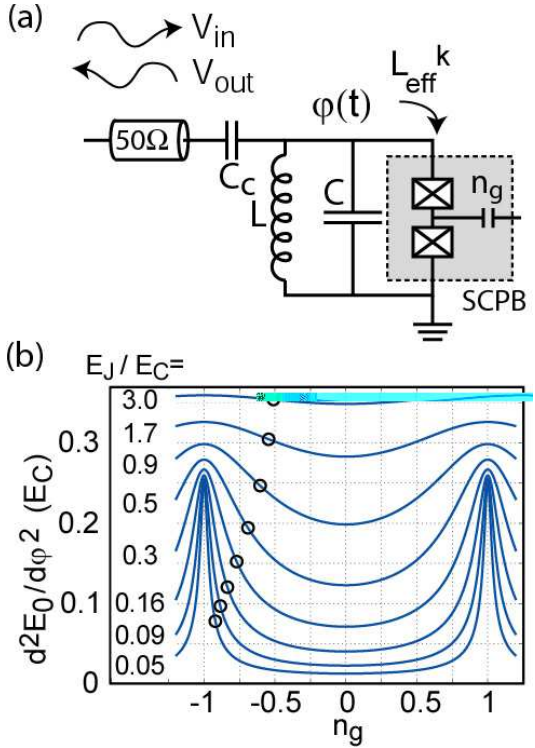}\label{fig:lset}
\caption{(a) Schematics of the capacitively coupled "L-SET"
inductive rf-electrometer. The resonance frequency $f_p^{-1} = 2
\pi \sqrt{(L \parallel L_{\mathrm{eff}}^k) C}$ depends on the SCPB
Josephson inductance $L_{\mathrm{eff}}^k$; (b) calculated
modulation of the second $n_g$ - derivative (inverse
$L_{\mathrm{eff}}^0$, see Eq.~(\ref{eq:d2EdI2})) at the SCPB
ground energy band, for different $E_J/E_C$, and $\varphi = 0$.
The circles mark optimal bias points for the electrometer
operation.}
\end{figure}

With respect to $\varphi$, the SCPT behaves as an inductance
(Fig.~\ref{fig:lset} (b)), dependent, first of all, on the band
index $k$, as well as on $n_g$ and $\varphi$:
\begin{equation}\label{eq:d2EdI2}
    L_{\mathrm{eff}}^{k} (n_g, \varphi) = \left( \frac{d^2E_k}{d \Phi^2}\right)^{-1} = \left(\frac{\Phi_0}{2 \pi} \right)^2 \left( \frac{d^2E_k}{d
\varphi^2}
    \right)^{-1}.
\end{equation}
The strong $n_g$ dependence of $L_{\mathrm{eff}}^0$ when $E_J/E_C
\ll 1$ has been used by the present authors to implement a fast
reactive electrometer \cite{lset}, using the scheme of
Fig.~\ref{fig:lset} (a). The measurements are performed by
studying the phase shift $\Theta =
\arg(V_{\mathrm{out}}/V_{\mathrm{in}})$ of the "carrier" microwave
reflected from a resonant circuit containing the SCPB. Denoting by
$Z$ the lumped-element impedance seen when looking towards the
resonance circuit from the transmission line of impedance $Z_0 =
50 \, \Omega$, the reflection coefficient of a voltage wave is
\begin{equation}\label{eq:gamma}
    \Gamma = \frac{V_{\mathrm{out}}}{V_{\mathrm{in}}} =
\frac{Z-Z_0}{Z+Z_0}= \Gamma_0 e^{i \Theta}.
\end{equation}
Since the whole setup consists in principle only of reactances,
the inductively read scheme should be superior in terms of noise
and back-action \cite{zorinrf} over the previous fast
electrometer, the rf-SET \cite{rfset}, which relies on the control
of dissipation.

\begin{figure}[h]
  \includegraphics[width=7.5cm]{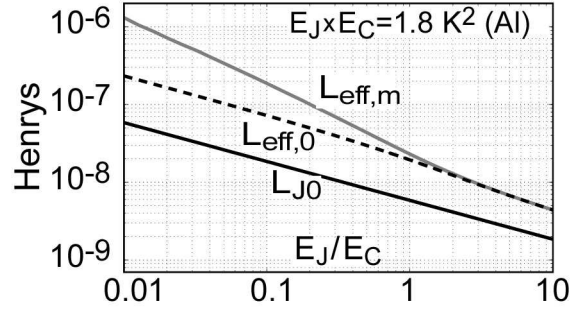}\label{fig:ljeff}
\caption{Numerical values of the SCPB ground band Josephson
inductance $L_{\mathrm{eff,0}}$ (that at $n_g = \pm 1, \varphi =
0$), and $L_{\mathrm{eff,m}}$ (at maximum gain, at $\varphi = 0$)
for a typical aluminium device. Also shown is the "classical"
Josephson inductance $L_{J0}$ in Eq.~(\ref{eq:LJNonLin}).}
\end{figure}

The crucial number for electrometer operation is the differential
modulation of $L_{\mathrm{eff}}$ (at the ground band), or
dimensionless "gain":
\begin{equation}\label{eq:g}
g \equiv \frac{\partial}{\partial n_g}
\left(\frac{L_{\mathrm{eff}}}{L_{\mathrm{eff,0}}} \right) \, ,
% = \frac{\partial}{\partial n_g} \left( \frac{\partial ^2
%E_0}{\partial \varphi^2} \right)^{-1} \left(\frac{\Phi_0}{2 \pi}
%\right)^2 \frac{1}{L_{\mathrm{eff,0}}}
\end{equation}
which we have presented as normalized by $L_{\mathrm{eff,0}}$
which denotes the Josephson inductance at the special point $(n_g
= \pm 1, \, \varphi = 0)$. Using Eq.~(\ref{eq:bands}), we have
$L_{\mathrm{eff,0}} = 4 L_{J0}$. For the best electrometer
performance, $n_g$ should be biased at the points marked by
circles in Fig.~\ref{fig:lset} (b). From Eq.~(\ref{eq:bands}) we
also find the maximum gain $g_m$ which grows rapidly when $E_J /
E_C \ll 1$: $g_m \simeq 2 (E_J / E_C)^{-1}$. Another important
figure is the value of $L_{\mathrm{eff}}$ at the optimal gate bias
which yields $g_m$, denoted here as $L_{\mathrm{eff,m}}$
\cite{AnalytNote}. To some extent, the rapidly growing
$L_{\mathrm{eff,m}}$ towards lowering $E_J/E_C$ (see
Fig.~\ref{fig:ljeff}) cancels the benefit of growing $g_m$ from
the point of view of charge sensitivity.

Without going into details, optimal charge sensitivity limited by
\emph{zero-point fluctuations} in the loaded $LC$-oscillator in
Fig.~\ref{fig:lset} (a) is \cite{lsetsensit}:
\begin{equation}\label{eq:sqFUND}
    s_q^{\mathrm{QL}} = \frac{16 \sqrt{2} e (L_{\mathrm{eff,m}})^2 \sqrt{2 k_B T_N} }
    {g_m \pi \sqrt{\hbar} \Phi_0 L_{J0} \sqrt{Q_i}},
\end{equation}
where $T_N$ is the noise temperature of the rf-amplifier, and
$Q_i$ is the \emph{internal} quality factor of the resonator.
Evaluating the values in Eq.~(\ref{eq:sqFUND}) numerically, we
find that $s_q \sim 10^{-7}$e$/\sqrt{\mathrm{Hz}}$, order of
magnitude better than the shot-noise limit of rf-SET, is
intrinsically possible for the L-SET if $Q_i \sim 10^3$ and $T_N
\sim 200$ mK. So far, the sensitivity in experiment
\cite{lsetsensit,MIKAthesis} has been limited by $Q_i \lesssim 20$
down to $s_q \simeq 2 \times 10^{-5}$e$/\sqrt{\mathrm{Hz}}$. The
limit of Eq.~(\ref{eq:sqFUND}) is reached when parameter values
are chosen so that
\begin{equation}\label{eq:omegaopt}
\omega_p = \frac{\Phi_0^2 (L_{\mathrm{eff,m}} + L)}{64 \hbar
L_{\mathrm{eff,m}} L}.
\end{equation}
Equation (\ref{eq:omegaopt}) yields values typically $f_p =
\omega_p/(2 \pi) \simeq 1-2$ GHz, though dependence of $f_p$ is
rather weak.

\section{Quantum capacitance}

The band energies of an SCPB depend on the (gate) charge $n_g$,
see Fig.~{\ref{fig:EnergySurface} (b), and the SCPB should then
behave like a capacitance with respect to changes of $n_g$
\cite{Likharev85,AverinBruder03}, which means that the point of
observation is at the gate electrode:
\begin{equation}
C_{\rm eff}^k = - \frac{\partial^2 E_k(\varphi, n_g)}{\partial
V_g^2} = - \frac{C_g^2}{e^2} \frac{\partial^2 E_k (\varphi,
n_g)}{\partial n_g^2}\,. \label{Ceff_defin}
\end{equation}
\emph{Phase} modulation of the input capacitance
$C_{\mathrm{eff}}(n_g, \varphi)$ of the SCPB observed in this
manner is plotted in Fig.~\ref{fig:QCap} (b). As seen in the
figure, $C_{\mathrm{eff}}$ has a strong phase dependence in the
limit $E_J/E_C \gg 1$ around $\varphi = \pm \pi$. Exactly at
$\varphi = \pm \pi$, Cooper-pair tunneling is completely blocked,
and $C_{\mathrm{eff}}$ reduces to classical series capacitance of
the junctions and $C_g$, that is, $\left[ (C_1 + C_2)^{-1} +
C_g^{-1} \right]^{-1}$.

The input capacitance depends sensitively (quadratically) on the
coupling capacitance $C_g$, and even when $C_g$ is made unusually
large such that it practically limits the charging energy,
$C_{\mathrm{eff}}$ typically remains very small, in the
femto-Farad range, see right hand scale of Fig.~\ref{fig:QCap}
(b). However, it has been suggested that the extremely strong
phase dependence could be used for fast, reactively read phase
detection \cite{Leif05}. This "CSET" mode of operation is somewhat
dual to the "L-SET" electrometry.

\begin{figure}[h]
  \includegraphics[width=7.5cm]{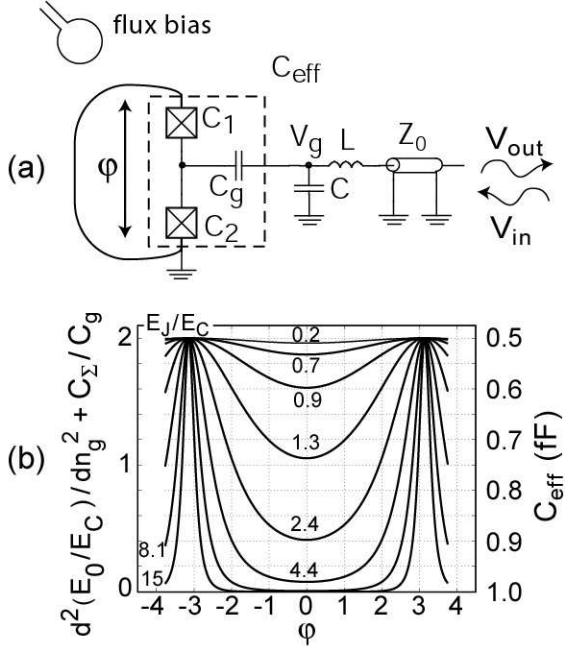}\label{fig:QCap}
\caption{(a) Schematics of the experiment used to study how the
SCPB appears as a tunable capacitance $C_\mathrm{eff}$; (b)
\emph{left scale} is the calculated second $n_g$-derivative of the
SCPB ground band at $n_g = 0$, and \emph{right scale} is the
corresponding effective capacitance if $C_g = 1$ fF and
$C_{\Sigma} = 2$ fF.}
\end{figure}

An important figure of merit for phase sensitivity is the
differential gain, analogous to Eq.~(\ref{eq:g}):
\begin{equation}\label{eq:gphase}
f \equiv \frac{\partial}{\partial \varphi}
\left(\frac{C_{\mathrm{eff}}}{C_g^2/(2C_{\Sigma})} \right) \, .
\end{equation}
The maximum of $f$ w.r.t. $\varphi$ at $n_g = 0$ is plotted in
Fig.~\ref{fig:PhaseSensit}.

We consider the experimental setup of Fig.~\ref{fig:QCap} (a),
where the quantum capacitance $C_{\mathrm{eff}}$ is in parallel
with a (generally much larger) stray capacitance $C$, and forms a
resonator with an inductance $L$. In this scheme, it is typical to
operate in the limit of vanishing internal dissipation which
corresponds to change of phase $\Theta$ of the reflected carrier
changing by $2\pi$ around the resonant frequency $f_p$.

Similarly as in the inductive readout, there are here no internal
noise sources except quantum fluctuations in the resonator.
Typically, therefore, sensitivity is again limited by noise of the
preamplifier: spectral density of the voltage noise referred to
preamplifier input is $s_{\mathrm{Vout}} = \sqrt{2 k_B T_N Z_0}$,
which can be regarded as a phase noise of the microwave carrier,
$s_{\Theta} = s_{\mathrm{Vout}} / V_{\mathrm{out}}$. When the
carrier amplitude is optimally large, it can be shown that under
the conditions mentioned, $V_{\mathrm{out}} = \frac{e}{2C_g} Z_0
\sqrt{\frac{C}{L} \frac{1}{2\pi}}$. When referred as an equivalent
flux noise at detector input using Eq.~(\ref{eq:gphase}), the
result becomes
\begin{equation}\label{eq:sthetaCSET}
\begin{split}
s_{\varphi} = & \frac{s_{\Theta}}{\partial \Theta / \partial
\varphi} =
    2\sqrt{\pi} e \left( \frac{C}{C_g} \right) \frac{\sqrt{k_B T_N Z_0}}{f_m E_{C}} \\
& \simeq \frac{4 \sqrt{\pi} C \sqrt{k_B T_N Z_0}}{f_m e} \, ,
\end{split}
\end{equation}
where the last form follows from the assumption that at high
$E_J/E_{C}$, charging energy is limited by the large gate
capacitance. This is the ultimate limit with advanced junction
fabrication (very thin oxide). The predicted phase sensitivity is
plotted in Fig.~\ref{fig:PhaseSensit}. Evidently, sensitivity
improves with decreasing stray capacitance $C$, since this results
in larger modulation of total capacitance $C + C_{\mathrm{eff}}$.
We see that $s_{\varphi} < 10^{-6}$rad$/\sqrt{\mathrm{Hz}}$, far
beyond an equally fast rf-SQUID, is possible in principle at high
$E_J / E_C \sim 10$ and a low stray capacitance $C \sim C_g$.

\begin{figure}[h]
  \includegraphics[width=7.5cm]{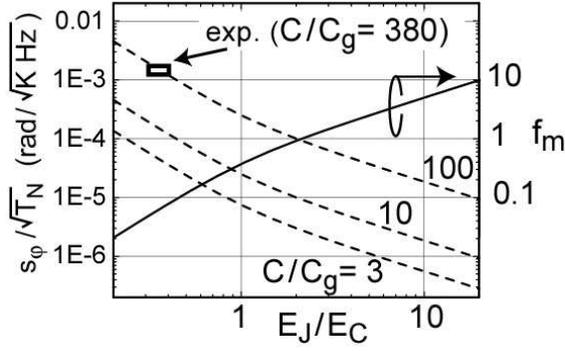}\label{fig:PhaseSensit}
\caption{Left scale (dashed lines): Phase sensitivity predicted
for the CSET, Eq.~(\ref{eq:sthetaCSET}), first form, if $Z_0 = 50
\, \Omega$ and $E_C = 1$ K, for different ratios of the gate
capacitance to stray capacitance. Right scale: the maximum gain
$f_m$ of the phase detector. Experimental point is given by the
rectangle (note that it had a larger capacitance ratio of $\sim
380$).}
\end{figure}

We investigated the discussed phase detection experimentally in
the scheme of Fig.~\ref{fig:QCap} (a), with the parameter values
$E_J = 0.30$ K, $E_C = 0.83$ K, $E_J/E_C = 0.36$, $C_g = 0.65$ fF,
$C = 250$ fF, $C/C_g = 380$, and $L = 160$ nH. Except $C_g$, the
sample parameters were determined by microwave spectroscopy
\cite{cap}. To the input bias coil of the phase detector, we
applied low-frequency modulation by $0.013 \, \Phi_0$ at 80 Hz.
Its amplitude was calibrated relying on $\Phi_0$-periodicity of
the static response. This way, we obtained a sensitivity of 1.3
mrad/$\sqrt{\textrm{Hz}}$, see the black curve in
Fig.~\ref{fig:FluxSensit}, limited by the 4 K amplifier noise,
which figure is even better than expected (see
Fig.~\ref{fig:PhaseSensit}).

We shall now discuss Fig.~\ref{fig:FluxSensit} in more detail.
Both the curves were measured at a flux bias close to $\varphi
\sim \pi$ which yields the largest gain $f_m$. For the black
curve, $f(n_g, \varphi)$ was further maximized by tuning $n_g$
close to 1, which also yielded a high level of low-frequency noise
as can be seen in the data. Since the low-frequency noise is
significantly reduced when we tuned $n_g = 0$ where the response
is insensitive to charge fluctuations (the gray curve), we assign
the increased noise around $n_g = 1$ to the ubiquitous
low-frequency background charge noise.

Since the low-frequency noise at $n_g = 0$ is free from the effect
of charge noise, we were able to directly measure in the scheme
the apparent flux noise, which we attribute to critical-current
fluctuations. The power spectrum of the gray curve shows $1/f^2$
dependence in contrast to typical $1/f$ rule \cite{Clarke04} for
big junctions. We convert this noise into fluctuations in critical
current of either of the junctions, in other words, we ask the
question: what would be the $I_C$ fluctuation $\Delta I_C =
2e/\hbar (\Delta E_J)$ in either one of the junctions which would
cause a capacitance fluctuation $\Delta C_{\mathrm{eff}}$, and
hence an apparent phase fluctuation $\Delta \varphi$ ? Equation
(\ref{eq:gphase}) implies
\begin{equation}
\Delta C_{\mathrm{eff}}(n_g, \varphi) = f(n_g, \varphi) C_0 \Delta
\varphi \, ,
\end{equation}
where we have marked $C_0 = C_g^2/(2C_{\Sigma})$. This then
converts into $E_J$ fluctuation according to
\begin{equation}\label{eq:deltaej}
\Delta E_J= \Delta C_{\mathrm{eff}} \left( \frac{\partial
C_{\mathrm{eff}}}{\partial E_J} \right)^{-1}
\end{equation}
We compute the partial derivative in Eq.~(\ref{eq:deltaej})
numerically; the result is $\frac{\partial
C_{\mathrm{eff}}}{\partial E_J} \simeq 0.072 \, \left(
\frac{C_g}{e} \right)^2 \frac{E_C}{E_J} \simeq 0.30 \left(
\frac{C_g}{e} \right)^2$. We also set $f(n_g, \varphi) \rightarrow
f_m$ since we had tuned to the maximum gain.

Finally, since the spectral densities of fluctuations are related
similarly as the fluctuations itself, we have the amplitude
spectrum of $I_C$ noise:
\begin{equation}\label{eq:deltaic}
s_{IC} =  \frac{2e}{\hbar} s_{EJ} = \frac{2e}{\hbar} f_m
s_{\varphi} C_0 \left( \frac{\partial C_{\mathrm{eff}}}{\partial
E_J} \right)^{-1} \, .
\end{equation}
This yields the gray line in Fig.~\ref{fig:FluxSensit}, with the
numbers around 10 Hz being comparable to big junctions.

\begin{figure}[h]
  \includegraphics[width=7.5cm]{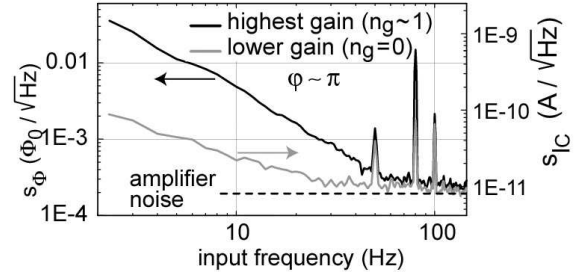}\label{fig:FluxSensit}
\caption{Measured equivalent flux noise at CSET input (left scale,
black curve) and critical current noise (right scale, gray curve).
Low-frequency flux modulalation by $0.013 \, \Phi_{0,
\mathrm{RMS}}$ at $80$ Hz was used as a marker.}
\end{figure}

\begin{theacknowledgments}

We thank T. Heikkil\"a, F. Hekking, R. Lindell, Yu. Makhlin, M.
Paalanen, and R. Schoelkopf for comments and useful criticism.
This work was supported by the Academy of Finland and by the
Vaisala Foundation of the Finnish Academy of Science and Letters.
\end{theacknowledgments}


\begin{thebibliography}{99}
\bibitem{AnalytNote} Equation~(\ref{eq:bands}) does not allow for an
analytical formula for $L_{\mathrm{eff,m}}$.
\end{thebibliography}
\end{document}